# A Modern-day Alchemy: Double Glow Plasma Surface Metallurgy Technology

Zhong Xu[1*], Jun Huang[2*], Hongyan Wu[3], Zaifeng Xu[1], Xiaoping Liu[1], Naiming Lin[1], Dongbo Wei[4], Pingze Zhang[4]

[1]Research Institute of Surface Engineering, Taiyuan University of Technology, Taiyuan 030024, China.
[2]School of Materials Science and Engineering, Nanchang Hangkong University, Nanchang 330063, China.
[3]Department of Material Physics，Nanjing University of Information Science and Technology, Nanjing 210044, China.
[4]College of Materials Science and Technology, Nanjing University of Aeronautics & Astronautics, Nanjing 211100, China.
[*]Corresponding author. Email: xuzhong@tyut.edu.cn; huangjun@nchu.edu.cn

**Abstract**：   In the long history of science and technology development, one goal is to diffuse solid alloy elements into the surface of steel materials to form surface alloys with excellent physical and chemical properties. On the basis of plasma nitriding technology, double glow plasma surface metallurgy technology has answered this challenge. This technology, which seems to be a modern-day alchemy, can use any element in the periodic table of chemical elements, including solid metal elements and their combinations, to form many types of surface alloyed layers with high hardness, wear resistance, corrosion resistance and high temperature oxidation resistance on various metal materials. For examples, nickel base alloys, stainless steels and high speed steels are formed on the surfaces of ordinary carbon steels; and high hardness, wear resistance and high temperature oxidation resistance alloy are formed on the surface of titanium alloy.

   This article briefly introduces the formation and principle of double glow plasma surface metallurgy technology, and summarizes the experimental results and industry application. The significance and development prospect of this technology are discussed.

## 1. Introduction

   Since ancient times, the people have dreamed of "turning stones into gold", and this dream has never been realized. An American reporter Mr. Beverly Phillips of the newspaper 《The State》 interviewed Professor Zhong Xu in July 1989. On August 6, 1989, the reporter published a long report in the newspaper. The title of the article was "Local firm marketing technology --- Xu-Loy system creates metal alloy on surfaces". The first sentence of this long report is "It may sound like modern-day alchemy, but it's not." "The process, a more efficient way of blending two metals to create one with enhanced surface properties, was developed by Zhong Xu." On July 24 in the same year, "Business Week" published an article titled "A BREAKTHROUGH IN MAKING METALS TOUGHER" in the column of Development to Watch by William D. Marbach. The article pointed out that "a Chinese scientist has developed a new surface-alloying method that produces alloys that are at least 40 times thicker than those created by conventional ion implantation." "The so-called Xu-loy process can create a stainless steel layer on cheaper carbon steel. It can also make high-strength surface alloys using metals such as Molybdenum and Tungsten."

The results of 40 years of experimental research have proved that the "Double Glow Plasma Surface Metallurgy Technology"，also called the Xu-Loy technology or the "Xu-Tec Process" is the contemporary alchemy that can "turn stones into gold".

## 2. Double Glow Plasma Surface Metallurgy Technology (Xu-Tec Process)

The Double glow plasma surface metallurgy technology was invented on the basis of the "plasma nitriding" technology invented by German scientist B. Berghaus in 1930 (1). Plasma nitriding technology is recognized by the world as one of the most important inventions in material surface technology. However, this technology can only be applied to non-metal elements such as nitrogen, carbon and sulfur. For nearly a century, people have tried to apply glow discharge to a large number of solid alloy elements with little success.

Based on many years of research on plasma nitriding technology, Professor Zhong Xu realized that the key problem is how to gasify and evaporate the solid alloy elements? so that the glow discharge atmosphere contains the atoms of alloy elements.

In a long-term observation of glow discharge phenomenon in the process of plasma nitriding, Xu observed the micro arc discharge phenomenon on the surface of the work-piece, and sometimes produced tiny "sparks" produced by ion bombardment sputtering of steel materials. In addition, an increasing amount of iron powder was observed in the furnace chassis. This made Xu realize that the iron powder deposited on chassis is precipitated from the gasification of solid iron elements by ion bombardment sputtering. Therefore, Xu decided to set up an intermediate electrode with negative potential between anode and cathode in plasma nitriding equipment. The intermediate electrode was made of the solid alloy elements which were intended to be diffused into work-piece to form surface alloy. In 1978, Xu found the double glow discharge phenomenon around the work-piece and the middle electrode respectively. On the basis of experimental study on the characteristics of double glow discharge, Xu invented "Double glow plasma surface metallurgy technology" in 1980.

Double glow plasma surface metallurgy is a kind of surface alloying technology, which uses double glow discharge phenomenon and low temperature plasma to form alloy layer with special physical and chemical properties on the surface of metal materials through ion bombardment, sputtering, space transmission, deposition and diffusion in low vacuum.

The basic principle of the Xu-Tec process is shown in Fig. 1. Three electrodes are set in a vacuum vessel: anode, sample cathode and source cathode. Two adjustable voltage DC power supplies are set between the anode and the work-piece and between the anode and the source, respectively. Both the work-piece and the source are in negative potential. When the chamber is vacuumed, a certain amount of argon gas is filled. Then, with the increase of voltages of the power supplies, glow discharges will be generated between the anode and the work-piece and between the anode and the source at the same time. This is the so-called "double glow discharge

phenomenon".(2) Under the condition of glow discharge, the positive ions of argon produced by glow discharge bombard the source electrode, which makes the alloying elements to be sputtered from the surface of the source electrode, and then transported and adsorbed on the surface of the work-piece through space. At the same time, the positive ion of argon bombards the work-piece to heat it to high temperature, and makes the alloy elements adsorbed on the surface of the work-piece to diffuse into the interior of the work-piece to form the surface alloy layer.

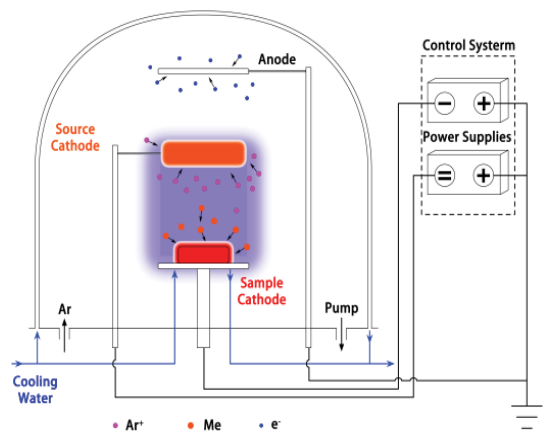

**Fig. 1 Principle of Xu-Tec Process**

The concept of Plasma "Surface Metallurgy" was first proposed by Professor Zhong Xu in 1982 at an international conference: "The 10th International Symposium on Discharge and Electrical Insulation in Vacuum" held in Columbia, South Carolina, USA (3).

In October 2017, the German Springer Publishing House published a monograph titled "Plasma Surface Metallurgy with Double Glow Discharge Technology--Xu-Tec Process", which introduced in detail "Double Glow Plasma Surface Metallurgy/Alloy Technology" (4).

The technology has been patented in the United States, Britain, Canada, Australia and Sweden (5-10).

## 3. Summary of experimental results

A large number of experimental results have shown that this technology can use any element in the periodic table of chemical elements, including solid metal elements and non-metal elements, to achieve surface alloying of metal materials. Countless kinds of surface alloys with excellent physical and chemical properties can be formed on the surface of steel materials, titanium and titanium alloys, copper and copper alloys, inter-metallic compounds and other conductive materials.

In order to improve the surface hardness, wear resistance, corrosion resistance and high temperature oxidation resistance of various metal materials, the following surface alloys have been prepared by using Xu-Tec process.

(1) Surface alloys formed on the surfaces of carbon steels:

Single element alloys：W, Mo, Al, Ti, Nb, Zr, Cu, Ta, V, Pt, Co, Au, Ni, Cr(11-14), etc.;

Multi-element alloys: W-Mo(2), Ni-Cr(15), Cr-Nb, Cr-Mo(16), Al-Cr(17), W-Mo-Ti, Co-W-Mo, Al-Cr-Nb, Ni-Cr-Mo-Cu, Ni-Cr-Mo-Nb(18, 19), Zr/ZrC(20), etc. Age-hardened high speed steels(19), stainless steels, Ni-based alloys and Super-alloys(18) etc..

As an example, using Inconel 625 alloy as sputtering target material, plasma surface alloying processes were conducted on surfaces of low carbon steel at 1000°C for 3 hrs.. Fig. 2 shows the micro-structure and distribution of Ni, Cr, Mo, Nb of the surface alloyed layer in industrial pure iron. The thickness of the alloyed layer is more than 40μm and contents of Ni, Cr, Mo, Nb are gradient decreased in sub-surface layers from the surface.

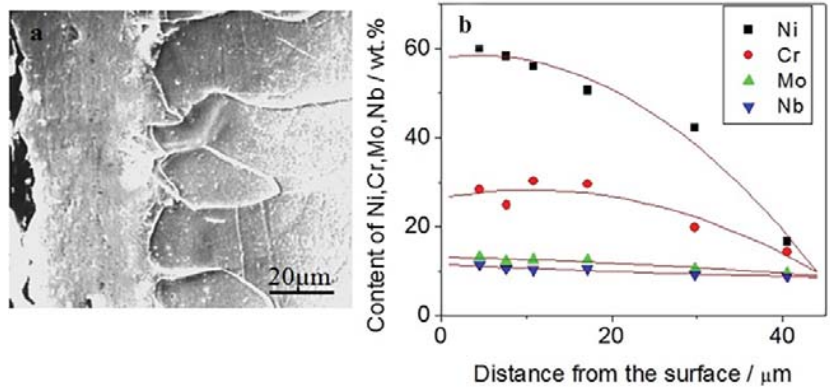

**Fig. 2 (a) Micro-structure and (b) Ni, Cr, Mo, Nb distribution of alloyed layer. (18)**

The above example illustrates two characteristics of surface alloying: (a) Alloying elements enter the metal matrix material through thermal diffusion to form a surface alloy; (b) The alloying elements in the surface alloy are distributed in gradient.

(2) Surface alloys formed on the surfaces of titanium alloys:
 Single element alloys: Mo, Cr, Cu, Zr, Nb, Ta, Pt, Pd(21-24);
 Multi-element alloys: Cr-Ni, Ti-Mo, Al-Y, Al-Cr, Mo-Ni(25), Zr-Y, Zr-Er(26), Cr-Co-Ni-Al-Ti-Y(27) and TiN, ZrC, ZrCN, TiSiN, etc. (28-30).
(3) Surface alloys formed on the surfaces of Ti-Al inter-metallic compounds:
 Single element alloys: Nb, W, Mo, Ta, Cr(31-34);
 Multi-element alloys: Ni-Cr, Cr-W(31), Cr-Mo(35), Cr-Nb(36), Nb-C(37), Zr-Y(38, 39), Ta-W(40), Ni-Cr-Mo-Nb(41) and gradient Cr/CrC(42), etc..
(4) Surface alloys formed on the surfaces of other metal materials
 W-Mo alloyed layer on iron-based powder metallurgy gear materials(43).
 Ti, Ni and Ta alloys on the surfaces of Cu and Cu-alloys(44).
 Ni-Co-Cr-Al-Y alloy on the surface of Inconel 718 alloy (45).
 Ir, Mo and Fe-Cr-Mo-Si alloys on the surfaces of Niobium alloy (46).
 Ir, Nb and Ir-Zr alloys on the surface of Mo (47, 48).
 Ta and W-Ta-V-Cr alloys on the surface of W.
 $CuZr_2$ alloy on the surface of Zr (49).
 Cr-Si alloy on the surface of Ta.
(5) Gradient ceramization of surface of metals to enhance adhesion between metal and ceramic.
 Gradient ceramics on surface of carbon steels：TiN, WC, TiC and Ti(CN) .
 Gradient ceramics on surface of titanium alloys：TaN, ZrC, ZrCN, TiSiN, etc..

(6) Metallization on surface of ceramic to realize welding between ceramic and metals.

Ti, Zr, W and other metal layers on the surface of TiSi30 and Si3N4 ceramics.

Ta，W，Mo，Ti，Ni，Cu alloys on the surface of diamond (50-52).

Mo, W, Ta, Fe-Al-Cr alloys on the surface of C/C composites (53-55).

Ir alloy on the surface of WC (56).

(7) Cu, Ag alloys formed on the surface of stainless steels to form antibacterial stainless steels (57).

In the past five years, more than 100 articles about Xu-Tec have been published (58). Recently, the Xu-Tec process has successfully formed functionalization layers, such as graphene-oxide film, nanocrystalline Nb layer, W/Ta multilayer, SiC/Ta$_x$C bilayer coating, Ta-modified layer on the surface of the metal alloys or semiconductors (e.g. quartz, sapphire), to apply in many fields of optical-electricity, high-temperature oxidation, superhard materials and so on.

Using graphene oxide film (GO) as the target material, Xu Tec process can directly form graphene film with good adhesion to the surface of a variety of matrix materials (59). This method can avoid the consumption of nickel and copper to prepare graphene. Xu Tec may be a promising method to prepare large-area graphene.

To expand the application of sapphire in electronic devices, the nanocrystalline Nb layer on sapphire was prepared by double glow plasma alloying technique on sapphire single crystalline substrate surfaces of (0001) substrate orientation (60). The Nb coating has a thickness of 600-700 nm, exhibiting elongated column-like grains structure. The as-deposited nanocrystalline Nb coatings exits (110) texture, and companies with (220) oriented grains.

A W/Ta multilayer was designed and deposited on the surface of pure copper by Xu-Tec technology (61), which aimed to provide the coating superior binding force and favorable matching of mechanical properties. The surface hardness of W/Ta multilayer coating is almost double that of the Cu substrate. Compared to the single W coating, the W/Ta multilayer coating does not peel during the wear process due to the existence of a transition interlayer (Ta layer). The favorable matching of mechanical properties between the W coating and Cu substrate is accomplished by the Ta transition interlayer. The wear rate of the multilayer coating at 500 ℃ is much lower than that of the Cu substrate with single W coating, which is only 1/65 of the substrate.

SiC/Ta$_x$C bilayer coatings were prepared on stainless steel surface by double glow plasma surface alloying technique to improve the wear and corrosion resistance of the substrate (62). It was found that the friction coefficient and specific wear rate of the coated specimens were significantly decreased than the untreated substrate. Meanwhile, the corrosion current density and $R_{ct}$ of the bilayer coatings was about 37.6% and 15 times of the stainless steel, respectively.

In order to enhance diamond nucleation and adhesion on cemented carbide substrates, the Ta diffusion layers with gradient structure were deposited by the combination with double glow plasma surface alloying technique and the reverse-sputtering process (63). The results of scratch testing indicated that the diamond coatings deposited on WC-Co substrates with Ta diffusion layers has higher adhesion due to the enhancement of cobalt diffusion barrier effect.

In conclusion, a large number of experimental results have shown that double glow plasma surface metallurgy technology is a very powerful surface alloying technology, which can transform low-grade materials into high-grade materials, and it should have a very broad industrial application prospect.

Double glow plasma surface metallurgy technology was awarded US "Energy-related Invention Program" (Project) and "Small Business Innovation Program" (Project) in 1989. In 1992, Xu won the National Technological Invention Award of China. In 1994, the Xu-Tec process was selected as one of 15 "Major Key Technology Project" by the Ministry of Science and Technology of China.

## 4. Example of Industry Application

Steel plate was our initial focus in industrial application research. A new "Double Glow Plasma Surface Metallurgy Ni-Cr alloy Corrosion Resistant Steel Plate" was formed by using "double glow plasma surface metallurgy technology". At the same time, the bimetal high speed steel saw blade was invented in the United States. We have successfully diffuse W and Mo alloying elements into the teeth of low carbon steel saw blade, forming an unprecedented "Double Glow Plasma Surface Metallurgy High Speed Steel Saw Blade". The saw blade is flexible, which meets the cutting performance standard of American bimetal HSS saw blade. Colloid mill (diameter 450 mm × 300 mm), was treated by W-Mo co-alloyed and carburizing, which increased its service life by five times.

Professor Hengde Li, a famous materials science scientist and former chairman of the International Union of Materials Research Societies (IUMRS), once pointed out that: "The formation of stainless steel on the surface of ordinary carbon steel is a dream of the world's metallurgical industry for long Times."

In 1984，the first industrial application of the Xu-Tec process was selected to form a Ni-Cr alloy layer on the surface of carbon steel plate. Our goal was to develop this new kind of plate to partly replace the conventional stainless steel plate.

We successfully prepared nickel-based alloy layer on the surface of low carbon steel plate with a size of 1 m long and 0.5 m wide. Fig. 3 shows the flange parts and steel plate after plasma surface metallurgy Ni-Cr alloying treatment. The corrosion resistance of the Ni-Cr alloyed steel plate is better than 316L stainless steel.

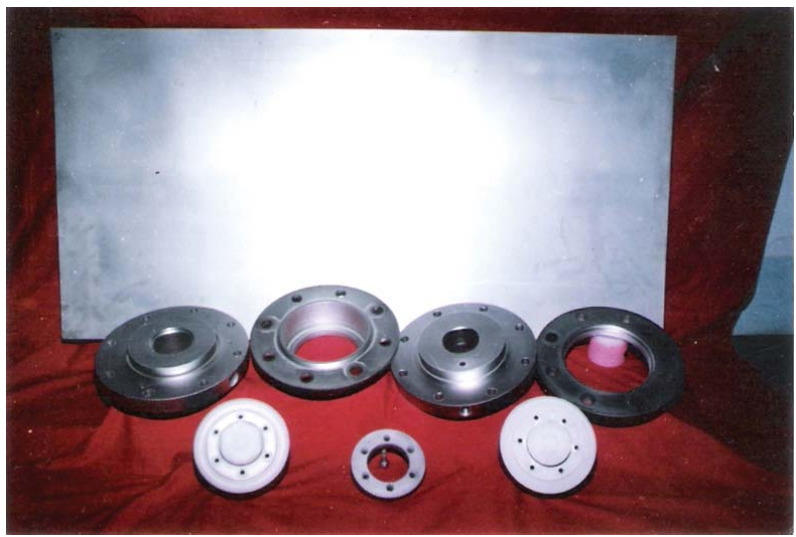

**Fig. 3 Steel plate and flanges after plasma surface alloying with Ni-Cr.**

The Plasma Surface Metallurgy Stainless Steel Plate may be the most important application of the Xu-Tec process.

The Xu-Tec process had been used to form Nickel base alloys with better corrosion resistance than 316L stainless steel on the surface of low carbon steel plate. According to a preliminary economic analysis, China in 2017 produced a total of 25.77 million tons of stainless steel, of which 85% is accounted for the stainless steel plate. If the plasma surface Ni-Cr alloying plate replaces 20% of stainless steel plates each year for the country, it is about 0.85 million tons of Nickel and 1.03 million tons of Chromium can be saved.

5. **Significance and Importance of Xu-Tec Process**

Agriculture, materials, machinery manufacturing and energy are the four pillar industries of human life.

Material is the matter basis of all science and engineering technology, and metal material is the most widely used material in machinery manufacturing industry. The parts in the machinery manufacturing industry are mostly made of steel, cast iron, titanium and titanium alloy, aluminum and aluminum alloy and other metal materials.

In the long-term production practice, people are more and more aware of the importance of material surface. The surface is just as important as the human skin, and it is a barrier to prevent external environmental damage (such as friction, impact of force, corrosion, oxidation, etc.).

A large number of facts have proved that the failure and damage of materials often start from the surface. In many working conditions, the requirements for the matrix and surface properties of materials are quite different. Generally speaking, the matrix is required to have high strength and certain toughness, while the surface is required to have high hardness, wear resistance, corrosion resistance, high temperature oxidation resistance, etc. In order to solve the contradiction between matrix and surface in performance requirements, many surface technologies have been invented to improve the hardness, wear resistance and corrosion resistance of material surface.

The existing surface technology can be basically divided into the following two categories: coating technology and surface alloying technology. Coating technology includes electroplating, deposition, coating, thermal spraying, physical vapor deposition, chemical vapor deposition and so on. It is characterized by mechanical combination between coating and substrate, but it is rather easy to peel and produce cracks under the working conditions of external force and temperature change. The characteristic of surface alloying is that the alloy elements enter the matrix through thermal diffusion and form the surface alloy layer. There is metallurgical bonding between the surface alloy layer and the matrix, forming a very good bonding strength.

Working conditions of mechanical parts (such as gears and shafts) are often affected by external forces, such as stretching, bending, impact, etc., as well as friction, wear, corrosion, temperature change resistance, etc. Therefore, for mechanical parts, surface alloying technology must be used to strengthen their surface properties.

In order to solve the contradiction between the performance requirements of metal parts surface and matrix, improve the surface hardness and wear resistance, carburizing and nitriding are the two most important surface alloying technologies. Solid alloy elements have been diffused into metal materials by solid, liquid and gas methods. Such as solid chromizing, liquid aluminizing, etc. However, these methods are very backward with poor working conditions, pollution, low efficiency, and can only be workshop type single piece small batch production. In recent years, the surface alloying technology of laser beam, ion beam and electron beam has been developed; but they have to be line scanning, and the equipment is expensive, energy consumption is large, efficiency is low, cost is high. Therefore, all of the above surface alloying technologies have shortcomings and limitations. It was a challenge to diffuse a large number of solid alloy elements into materials to form surface alloys with special physical and chemical properties.

Xu-Tec can use any chemical elements, including solid alloy elements and non-metallic elements and their combination to form numerous kinds of alloys with excellent physical and chemical properties on the surface of conductive metal materials to meet the needs of mechanical manufacturing products.

Although this technology was invented 40 years ago, it is still the most advanced, practical and powerful surface alloying technology in the world. This technology will have a broad application prospect in machinery manufacturing, transportation, household appliances, marine engineering, aerospace and other fields.

Throughout human history, due to the extreme importance of materials, historians use materials as a symbol of dividing human ages, such as the Stone Age, Pottery Age, Bronze Age, Iron Age, etc. Nowadays, computer technology based on functional materials has produced new industries of wireless communication, the Internet and artificial intelligence. However, all these new industries are still inseparable from metal materials and machinery manufacturing.

The Xu-Tec Process is a major breakthrough in the surface alloying technology, and an original major invention in the field of metal materials and mechanical manufacturing. This technology can greatly improve the performance of metal materials, comprehensively improve the quality and life of machine-made products, save a lot of precious alloy elements for humans, and create huge economic benefits, benefit of all mankind.

## 6. Conclusion and Prospect

(1) How to realize plasma surface alloying/metallurgy of metal materials by solid metal elements to form surface alloy with special physical and chemical properties is not well solved problem in a long time of human history.

(2) The Xu-Tec process，invented by plasma nitriding technology and double glow discharge phenomenon，can use any solid metal element and its combination to form countless surface alloys on the surface of conductive materials. This is indeed a modern-day alchemy and has opened up a new field of "Plasma Surface Metallurgy". This technology will be an important trend and direction for the future of metal surface engineering technology.

(3) The Xu-Tec process is a typical physical process. Its entire process includes the vacuum, glow discharge, ion bombardment, sputtering, deposition, diffusion, etc. There is no pollution. In addition, it consumes very little alloy elements since the surface alloy layer thickness is only about 0.1-1% of the thickness of the work-piece. This technology is a truly resource-saving and environment-friendly.

(4) The plasma Surface Metallurgy Technology is different from Plating, Coating, Deposition Spray, PVD, CVD and Thin film technology. The composition of the surface alloy layer has gradient distribution between surface alloy and matrix material. Therefore, the bond between the surface alloy and the matrix is very strong.

（5）The Xu-Tec Process has been successfully applied to carbon steel plate, cutting tools, colloid mill, chemical valves and other industrial products for special purposes. This technology transforms low-grade materials into high-quality materials, hence there should be countless industrial applications with significant economic impact.

(6) Double glow plasma surface metallurgy technology is an interdisciplinary technology involving vacuum and gas discharge, sputtering and deposition, diffusion and phase transformation, surface physics and chemistry, electrical power supplies and computer control. The mechanism of sputtering and diffusion under ion bombardment, the interaction between alloying elements and matrix elements, the control of composition of surface alloy，the stability of glow discharge and the suppression of arc discharge are all subjects that need to be further optimized. The solution of these challenging problems will further release the potential power of this technology.

The discovery of double glow discharge phenomenon and the invention of double glow plasma surface metallurgy technology fully prove that：Experimental observation is the source of invention and creation；Physics is the basis of invention and creation; There is still a lot of room for invention in classical physics.

## Acknowledgment

The authors wish to thank their colleagues and all doctorate students for their contributions in the Xu-Tec process.

**Fig. 1 Principle of Xu-Tec Process**

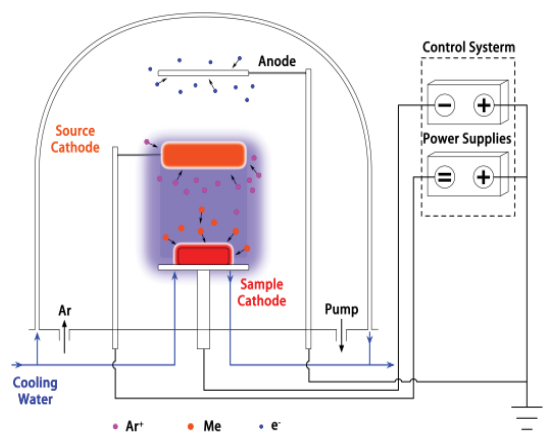

**Fig. 2 (a) Micro-structure and (b) Ni, Cr, Mo, Nb distribution of alloyed layer.**

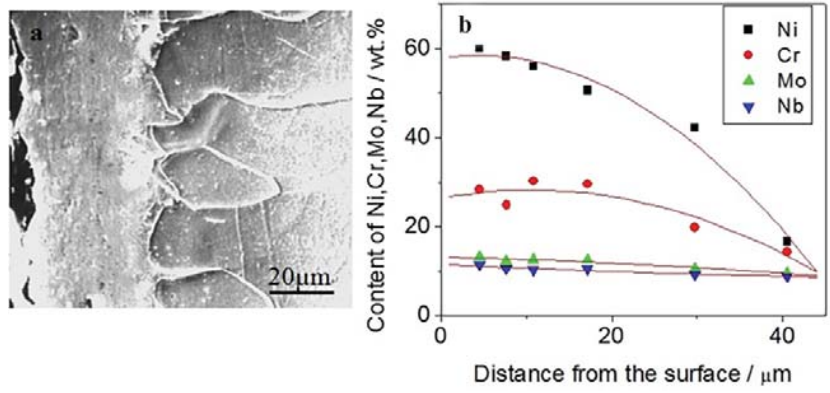

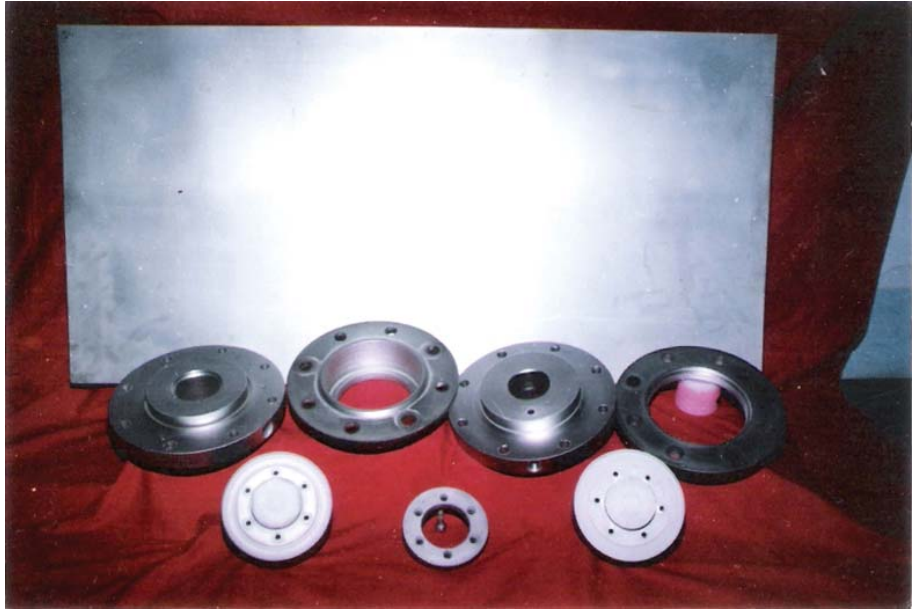

**Fig. 3 Steel plate and flanges after plasma surface alloying with Ni-Cr.**